\documentclass[aps,prd, twocolumn,showpacs,superscriptaddress]{revtex4}
\usepackage{amssymb,amsmath,graphicx,color,microtype}
\usepackage{graphicx}
\usepackage{txfonts}
\usepackage{epsf}
\usepackage{epstopdf}
\usepackage {amssymb}
\usepackage{multirow}
\newcommand{\nc}{\newcommand}
\nc{\ba}{\begin{eqnarray}}
\nc{\ea}{\end{eqnarray}}
\newcommand\be{\begin{equation}}
\newcommand\ee{\end{equation}}

\nc{\e}{{\bf{e}}}
\nc{\kk}{{\bf{k}}}
\nc{\pp}{{\bf{p}}}

\nc{\bfk}{{\bf{k}}}
\nc{\bfx}{{\bf{x}}}
\nc{\bfp}{{\bf{p}}}

\nc{\eH}{{\epsilon_H}}
\nc{\calP}{{\cal P}}
\nc{\im}{{ \mathrm{Im} } }

\begin{document}
%%%%%%%%%%%%%%%%%%%%%%%%%%%%%%%%%%%%%%%%%%%%%%%%%%%%%
	
\title{A Soliton Solution for the Central Dark Masses in  47-Tuc Globular Cluster and Implications for the Axiverse}

\author{Razieh~ Emami}
\email{razieh.emami_meibody@cfa.harvard.edu}
\affiliation{Center for Astrophysics, Harvard-Smithsonian, 60 Garden Street, Cambridge, MA 02138, USA}
\affiliation{ Institute for Advanced Study, The Hong Kong University of Science and Technology, Clear Water Bay, Kowloon, Hong Kong}

\author{Tom~Broadhurst}
\affiliation{Department of Theoretical Physics, University of the Basque Country UPV-EHU, 48040 Bilbao, Spain}
\affiliation{IKERBASQUE, Basque Foundation for Science, Alameda Urquijo, 36-5 48008 Bilbao, Spain}
\affiliation{DIPC, Basque Country UPV/EHU, E-48080 San Sebastian, Spain }

\author{George~ Smoot}
\affiliation{ Helmut and Anna Pao Sohmen Professor-at-Large, IAS, Hong Kong University of Science and Technology,
Clear Water Bay, Kowloon, 999077 Hong Kong, China}
\affiliation{Paris Centre for Cosmological Physics, APC, AstroParticule et Cosmologie, Universit\'{e} Paris Diderot,
CNRS/IN2P3, CEA/lrfu, Observatoire de Paris, Universit\'{e} Sorbonne Paris Cit\'{e}, 10, rue Alice Domon et Leonie Duquet,
75205 Paris CEDEX 13, France}
\affiliation{Physics Department and Lawrence Berkeley National Laboratory, University of California, Berkeley,
94720 CA, USA }

\author{Tzihong~Chiueh}
\affiliation{Department of Physics, National Taiwan University, 10617, Taipei, Taiwan}
\affiliation{Institute of Astrophysics, National Taiwan University. 10617, Taipei, Taiwan}
\affiliation{Center for Theoretical Physics, National Taiwan University, 10617, Taipei, Taiwan}

\author{Hoang~Nhan ~Luu}
\affiliation{ Institute for Advanced Study, The Hong Kong University of Science and Technology, Clear Water Bay, Kowloon, Hong Kong}

\begin{abstract}
We offer a standing wave explanation for the rising proper motions of stars at the center of the globular cluster 47-Tucanae, amounting to $\simeq 0.44\%$ of the total mass. We show this can be 
explained as a solitonic core of dark matter composed of light bosons, $ m \geq 10^{-18} eV $, corresponding to $ \leq 0.27 pc$, as an alternative to a single black hole (BH) or a concentration of stellar BH remnants proposed recently. This is particularly important as having a concentrated stellar BH remnant with the above radii is very challenging without the heavy core since the three body encounters would prevent the BHs to be that concentrated.  We propose this core develops from dark matter captured in the deep gravitational potential of this globular cluster as it orbits the dark halo of our galaxy. This boson may be evidence for a second light axion, additional to a  lighter boson of $10^{-22} eV$, favored for the dominant dark matter implied by the large dark cores of dwarf spheroidal galaxies. The identification of two such light bosonic mass scales favors the generic string theory prediction of a wide, discrete mass spectrum of axionic scalar fields. 

\end{abstract}

\maketitle

\section{Introduction}
Light scalar fields are a compelling choice for extending the standard model of particle physics, naturally generating axion-like dark matter with symmetry broken by the simple misalignment mechanism \cite{Preskill:1982cy,Abbott:1982af,Dine:1982ah,Khlopov:1985jw}. Such fields are generic to string theory from the dynamical compactification to 4 space-time dimensions describing our Universe \cite{Svrcek:2006yi}. These axion modes are expected to start out massless for symmetry reasons, subsequently picking up a relatively small mass by non-perturbative tunneling that is typically exponentially suppressed \cite{Arvanitaki:2009fg,Acharya:2010zx, Cicoli:2012sz, Hui:2016ltb}, resulting in a discrete mass spectrum of independent axions spanning many orders of magnitude.

Each axion field can develop rich structure on the de-Broglie scale \cite{Schive2014,Schive:2014hza, Veltmaat:2018dfz,Ringwald} under gravity, summed over the ensemble of these independent axion fields, which has been shown to account for the observed coldness of dark matter and the puzzling properties of dwarf galaxies for a dominant scalar field of $10^{-22}eV$ \cite{Schive2014,Schive:2014hza, Veltmaat:2018dfz,Ringwald}. Most conspicuously, a prominent soliton develops quickly at the center of every bound halo, as identified in the first simulations in this context \cite{Schive2014,Schive:2014hza, Mocz:2017wlg}. These solitons represent the ground state where self-gravity is balanced by an effective pressure arising from the Uncertainty Principle, yielding a static, centrally located and highly nonlinear density peak, or soliton. The soliton scale depends on the gravitational potential depth and for the favored $10^{-22}eV$ dominant dark matter this is predicted to be $\simeq 150pc$ for the Milky Way \cite{Schive2014,Schive:2014hza, Chen:2016unw}, much smaller than the size of the galaxy. This field may be detected directly by its inherent Compton scale pressure oscillation, at frequency $2m$ \cite{Khmelnitsky:2013lxt}. This is feasible using pulsars near the Galactic center for which a sizable 200ns timing residual is predicted on a convenient $~2$ months timescale that is boosted in amplitude within the relatively high dark matter density within the central soliton \cite{DeMartino:2017qsa}. 
	
In addition to this $10^{-22}eV$ axion for the dominant dark matter, a lighter axion of $10^{-33}eV$ may be considered to provide the dynamical dark energy from the associated quantum pressure \cite{Kamionkowski:2014zda,Emami:2016mrt,Poulin:2018dzj}, or as a related, probabilistic consequence of the string landscape \cite{Tye:2016jzi}. 

Axions that are heavier than $10^{-22}eV$ may also be anticipated, with sub-dominant but possibly significant contributions to the total dark matter density.  One possible place to look for their presence is at the center of globular clusters (GC) where a compact dark mass of over few thousands of solar mass is expected, although its origin is still under debait. Here we consider a very well studied GC  47-Tuc resolved very recently in Ref. \cite{Mann:2018xkm}. 

In this paper, we propose that the expected compact dark mass might be a soliton which is arisen from axion with some certain mass. Due to the compactness of this structure the dominant dark matter axion with mass of order $10^{-22} eV$ may not play any roles. Here we aim to estimate the parameters of such a soliton structure and to estimate the axion mass. 

We derive the parameters of this soliton from the first principles and add the contribution from the luminous matter as well. 

The paper is organized as the following. In Sec. \ref{sol-M-R-Relation}, we derive the mass-radius relation for a soliton profile. In Sec. \ref{proj-vel-sol} we compute the impact of the soliton in the projected velocity dispersion and 
fit the soliton model with the most recent observational data of 47-Tuc. We find the best fit values for the soliton mass and radius and find the associated axion mass. In sec. \ref{fit-imbh} we consider the fit for the case of a compact object. We conclude in Sec. \ref{conc}. 

\section{Soliton Mass-Radius Relation}
\label{sol-M-R-Relation}
In this section, we derive the solition mass-radius relation from the first principles. We follow the standard approach in Ref.  \cite{Hui:2016ltb, Chavanis:2011zi} for mapping the Bose Einstein Condensate (BEC) system to a hydrodynamical system, 
\ba  
\label{Euler}
&& \frac{\partial u}{\partial t} + (u \cdot \nabla)u = - \nabla \Phi - \frac{1}{m}\nabla Q,
\ea
here $Q$ denotes the quantum potential and it is given by,
\ba
\label{quantum-potential}
Q &\equiv& - \frac{\hbar^2}{2m}\frac{\Delta \sqrt{\rho}}{\sqrt{\rho}}
= -\frac{\hbar^2}{4m} \bigg{[}\frac{\Delta \rho}{\rho} - \frac{1}{2} \frac{(\nabla \rho)^2}{\rho^2} \bigg{]},
\ea
The total energy of this system is a summation of the kinetic term (both classical and quantum) and the potential term and is given by, 
\ba
\label{total-E}
E_{tot} &=& \frac{3}{4} M \left(\frac{dR_s}{dt}\right)^2 + \frac{3}{4} \frac{\hbar^2}{m^2}\frac{M}{R_s^2} - \frac{1}{\sqrt{2\pi}}\frac{G M^2}{R_s} \nonumber\\
&=& \frac{3}{4} M \left(\frac{dR_s}{dt}\right)^2 + V(R_s),
\ea

Where we have assumed a Gaussian profile for the density profile of soliton, $\rho_s(r) = M \left(\frac{1}{\pi R_s^2}\right)^{3/2} \e^{-r^2/R_s^2}$. It can be shown that the results from this consideration is in good agreement with the numerical calculation, \cite{Hu:2000ke,Schive2014, Schive:2014hza, Hui:2016ltb}.

$V(R)$ is the effective potential of the system and is given by $V(R) = \frac{3}{4} \frac{\hbar^2}{m^2}\frac{M}{R^2} - \frac{1}{\sqrt{2\pi}}\frac{G M^2}{R}$ where hereafter we drop the \textbf{sub-index $s$} from all of the quantities except in the density  and will return it at the end of the calculations. This is required in our variation calculation. 
Adopting a similar technique that determines the Chandrasekhar mass, the stable, time independent 
soliton core of the system is found by looking at the critical point of the  effective potential and by neglecting $dR/dt=0$. This is equivalent with the virial condition and gives us, 
\ba
\label{mass-radius-relation}
M(R) = \left(\frac{3\sqrt{2\pi}}{2}\right) \left(\frac{\hbar^2}{m^2 G R}\right),
\ea
which is identified as the  minimum since the second derivative of the effective potential is positive, $V''(R) = \frac{1}{\sqrt{2\pi}} \frac{G M^2}{R^3} >0$.

Next, we compute the back-reaction of the luminous matter on the above soliton mass-radius relation. 

We start with presenting the gravitational potential for a distribution of matter, 
\ba
\label{Gravitational-pot}
\Phi_L(r) 
&& = -\left(\frac{4 \pi G}{r}\right) \int_{0}^{r} r'^2 \rho_L(r') dr' - \left(4 \pi G\right) \int_{r}^{\infty} r' \rho_L(r') dr', \nonumber\\
\ea
where $\rho_L(r)$ denotes the luminous matter density for which we consider the King model \cite{Mann:2018xkm}, 
\ba 
\label{King_model}
\rho_{L}(r) =  K \left[ \frac{1}{\left( a^2 + r^2 \right)^{3/2}} - \frac{1}{\left( a^2 + r_t^2 \right)^{3/2}} \right], ~~~~(r \leq r_t), 
\ea
where $a$ refers to the effective core radius and $r_t$ denotes the tidal radius of King model. It also scales with a constant $K$. 

Finally the interacting gravitational energy for the system  includes the interaction between soliton-luminous, luminous-soliton and luminous-luminous is given by,
\ba
\label{gravitational-energy}
&W(R)&= \frac{1}{2} \int_{0}^{\infty} \rho_{s}(r, R) \Phi_{L}(r) d\vec{r} + \frac{1}{2} \int_{0}^{\infty} \rho_{L}(r, R) \Phi_{s}(r) d\vec{r} \nonumber\\
&&+ \frac{1}{2} \int_{0}^{\infty} \rho_{L}(r, R) \Phi_{L}(r) d\vec{r} = W_1(R) + W_2(R) + W_3(R), \nonumber\\
\ea
We notice here that while all of the integrals go to infinity, in practice they are truncated at $r = r_t$ which is the cutting threshold in the King model. 

We estimate the interaction gravitational energy as, 
\ba 
\label{king-interaction-total}
&&W(R) =  -4 \pi^3 G a^5 K^2 -\sqrt{16 \pi} \left(\frac{a^3}{R} G M K \right) \times \nonumber\\ 
&&  e^{\left(a/\sqrt{2}R \right)^2}BesselK\left(0, \left(\frac{a}{\sqrt{2}R}\right)^2 \right),
 \ea
where the first term comes from the luminous-luminous interaction and the second term is due to the interaction between the luminous and soliton terms.

We next compute the derivative of the interaction gravitational energy as $dW(R)/dR |_{R = R_s}  = 2 \pi G K M R_s$. Since these are evaluated at $R = R_s$ from now on we return to $R_s$ notation.

Finally we add this term to the pure contribution from the soliton itself, given by $V(R_s) = \frac{3}{4} \frac{\hbar^2}{m^2}\frac{M}{R_s^2} - \frac{1}{\sqrt{2\pi}}\frac{G M^2}{R_s}$, where as mentioned above we have replaced everywhere $R$ with $R_s$, 
and infer the mass-radius relation as, 
\ba  
\label{mass-radius-king}
M_s &\equiv&   M(R_s,m,a)  \nonumber\\
 &=&  -\left( \frac{3}{2 G}\sqrt{2\pi}\right)\left[
3 \sigma_L^2 \left(\frac{R^3_s}{a^2}\right) +  \left(\frac{\hbar^2}{m^2 R_s}\right) \right].
\ea
where for brevity we defined $\sigma^2_L \equiv  G M_{GC} / (2 R_{GC})$.

\section{Inferring the Soliton parameters from Velocity Dispersion}
\label{proj-vel-sol}
Having presented the soliton mass-radius relation, here we calculate the impact of the soliton in the projected velocity dispersion and estimate the parameters of soliton model including the soliton mass and radius as well as the associated axion mass.

\subsection{Projected velocity dispersion}

\textbf{First, we compute the velocity dispersion as, }
\ba 
\label{vel-dispersion}
\sigma^2(r) =  \frac{G}{\rho_L(r)} \int_{r}^{\infty} \frac{\rho_L(r') M(r')}{r'^2} dr,
\ea
where $\rho_L(r)$  refers to the King model and $M(r) = 4 \pi \int_{0}^{r} \left( \rho_L(r') + \rho_s(r') \right) r'^2 dr'$ denotes the interior mass to a sphere of radius $r$. Since the luminous and soliton contributions are separable, in the following we compute the soliton effect and add the luminous part to this from the recent observations. 

Next, we compute the projected velocity dispersion as given by  \cite{Mann:2018xkm},
\ba 
\label{Projected-Velocity}
\sigma^2_p(R) = \frac{\int_{R}^{\infty} \sigma^2(r) \rho_L(r) r \left(r^2 - R^2 \right)^{-1/2} dr }{\int_{R}^{\infty}  \rho_L(r) r \left(r^2 - R^2 \right)^{-1/2} dr}.
\ea

\subsection{Inferring the soliton parameters}

Having presented the projected velocity dispersion, here we estimate the soliton mass and radius from a direct comparison with the most recent observations of 47-Tuc as performed by \cite{Mann:2018xkm}.
In order to compare our results with that of Ref. \cite{Mann:2018xkm}, we choose a King model as our stellar profile and we fix the effective core radius ($a = 43.7$ \rm{arcsec}) and tidal radius ($r_{t} = 42$ \rm{arcmin})  to be 
the best fit given in King model. 
We take the distance of 47-Tuc from the earth ($D = 4.69$ \rm{kpc}). 
We add few different stellar contributions to the above soliton effect. They include the  effect from Stars, Binaries, White Dwarfs and the Neutron Stars. They are all presented in figure 6 in Ref. \cite{Mann:2018xkm}.

We preform a Markov Chain Monte Carlo (MCMC) analysis using the publicly available code \textit{emcee} \cite{ForemanMackey:2012ig} to estimate the value of soliton mass and radius. We make a two dimensional grid for these two parameters and for every points in this grid, we compare the projected velocity dispersion with the most recent observations. We assume a Gaussian Likelihood with uninformative priors for the soliton mass and the logarithm of soliton radius selected in the ranges $ 500 \leq M_{s}/M_{\odot} \leq 3 \times 10^4$ and $ 10^{-3} \leq R_{s}/pc \leq 10$, respectively. The lower limit in our sampling of the soliton radius is coming from the resolution limit of the experimental data which we take it to be $10^{-3}$ \rm{pc}. In our analysis, we use 300 walkers and run MCMC for $10^4$ steps in total. We monitor the time series of soliton mass and radius in the chain. We remove the first 700 steps to get a flat list of the samples.

\begin{figure}[!h]
\centering
\includegraphics[width=0.5\textwidth]{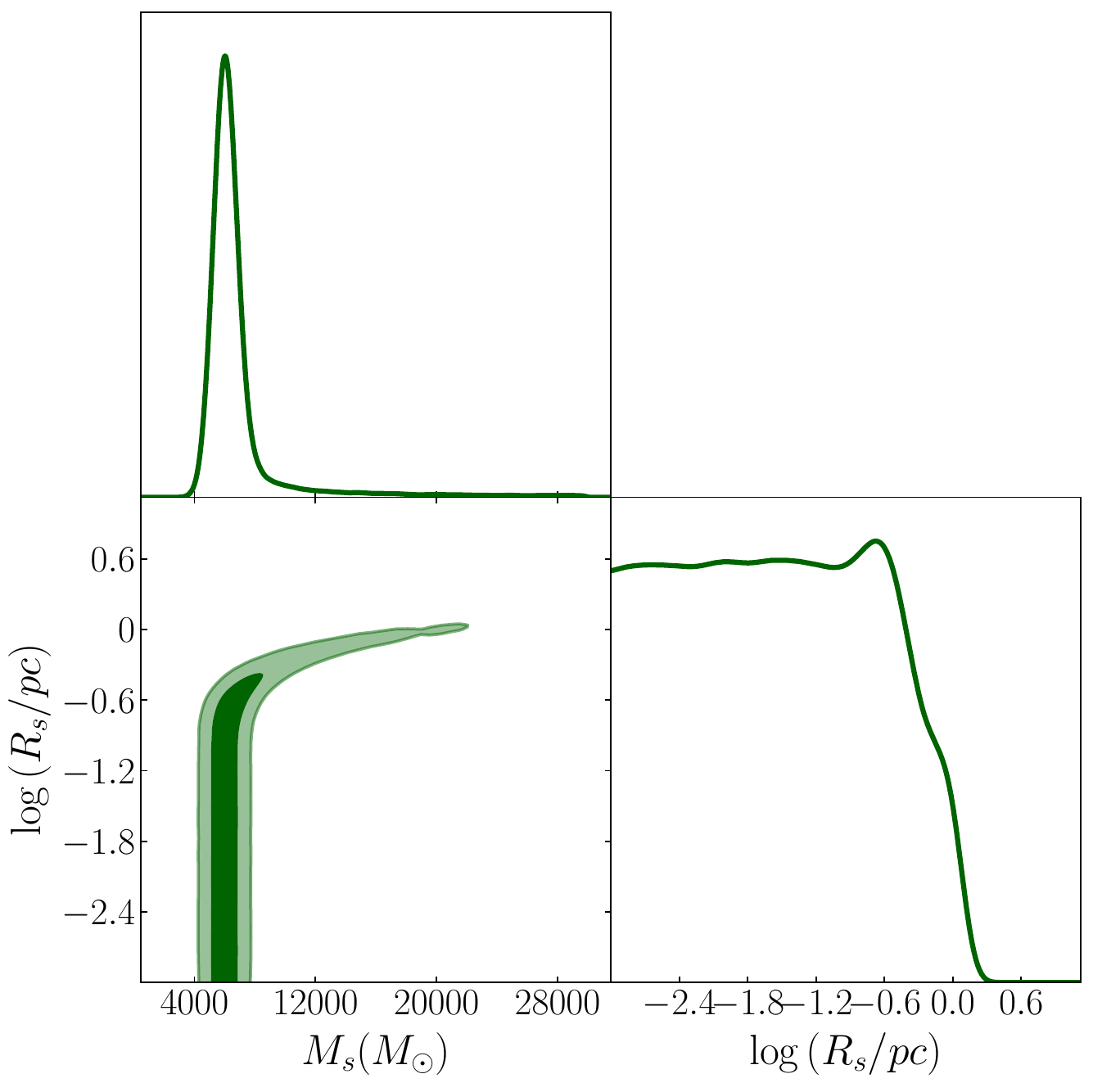}
\caption{The posterior of the soliton mass and logarithm of its radii as well as the allowed range for the soliton mass and radius. We use the projected velocity dispersion for the 47-Tuc GC. The vertical axes is cut off for $R_s \leq 10^{-3} \rm{pc}$ due to the resolution limitation. Here the contours show 68\% and 95\% confidence levels, respectively. } 
\label{mass-radius-sol}
\end{figure}

Figure \ref{mass-radius-sol} presents the resulting posterior distribution of the soliton mass and logarithm of its radii as well as the allowed range for the soliton mass and radius.

%%%%%%%%%%%%%%%%%%%%%%%%%%%%%%%%%%%%%%%%%%%%%%%%%
%%%%%%%%%%%%%%%%%%%%%%%%%%%%%%%%%%%%%%%%%%%%%%%%%
From this we achieve the following Mass and Radius, 
\ba 
\label{Sol-M-R}
M_s  &=&  6189.9^{+1200.89 }_{-846.41} M_{\odot} , \\
\label{Sol-M-R-1}
R_s & \leq &  0.27 pc.
\ea
where we have used the standard percentile method to compute the confidence levels.

\begin{figure}[!h]
\centering
\includegraphics[width=0.5\textwidth]{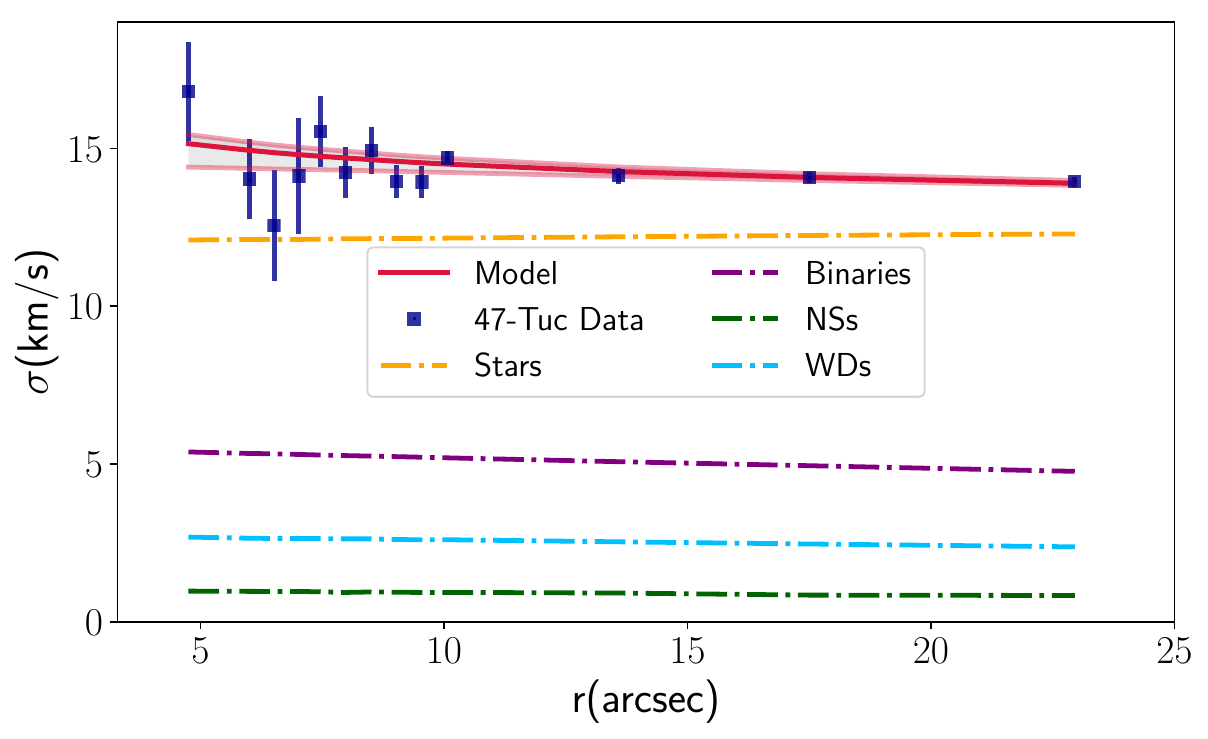}
\caption{Allowed values of the soliton mass and radius using the projected velocity dispersion for the 47-Tuc GC.  Theoretical model includes the effect of the soliton as well as the  Stars, Binaries, Neutron Stars (NSs)  and the White Dwarfs (WDs) as presented in figure 6 in Ref. \cite{Mann:2018xkm}.}
\label{velocity-projected}
\end{figure}

As a consistency check, in  figure \ref{velocity-projected} we use the above inferred range for the soliton mass and radius and compute the projected velocity dispersion and plot them against the data from Ref. \cite{Mann:2018xkm}. 

Our soliton profile is concentrated within a projected radius of  $\leq 0.27 \rm{pc}$. This is consistent with Mann et. al. (Ref. \cite{Mann:2018xkm}) who concluded ( at page 9, first column) that a mass 
concentration of stellar BHs with a scale of $\simeq 2 \rm{arcsec} = 0.045 \rm{pc}$ and a total mass of 19000 $M_{\odot}$ may also explain the rising central velocity dispersion of the visible stars. 
This is a much smaller than the core radius of the visible stellar light and is hence attributed to collection of  $\sim 1900$ stellar black holes (with the mass of order 10 $M_{\odot}$) that they assume has become
very concentrated relative to the stars under dynamical friction. This maybe very challenging as the ejection of such objects by three body encounters that may significantly deplete GC cores of such relatively massive compact remnants.

%%%%%%%%%%%%%%%%%%%%%%%%%%%%%%%%%%%%%%%%%%%%%%%%%
%%%%%%%%%%%%%%%%%%%%%%%%%%%%%%%%%%%%%%%%%%%%%%%%%
Finally we plug the above constraints for the soliton Mass and Radius back in the mass-radius relation and estimate the axion mass as, 

\ba  
\label{Axion-mass}
m \geq  1.07 \times 10^{-18} eV.
\ea

In addition to the above practical limits, the absence of evidence for a BH from Radio/Xray deep imaging can be also translated in a conceptual cut off in the value of $R_s$ to be bigger than the Schwarzschild radius associated with a soliton with mass $M_s \simeq  6189.9   M_{\odot}$. This gives us an upper limit on the mass of the axion to be $m \leqslant 2.7  \times 10^{-14} eV$. 

\section{Comparison with a single compact object}
\label{fit-imbh}
So far we have been only focused on the case of soliton as an alternative to IMBH motivated by the lack of gas accretion in the latter scenario. We have shown that our soliton should be relatively compact to be consistent with the observations. It is therefore motivated to find the actual fit for the case of a compact object as well. This enables us to see how much the observations of the velocity dispersion alone could potentially distinguish among these two scenario. 

For this purpose we replace $M(r')$ in Eq. (\ref{vel-dispersion}) with the mass of compact object and compute the projected velocity dispersion. Then we perform our MCMC analysis for this case and find the mass of the compact object. In our analysis, we use 300 walkers and run MCMC for $700$ steps in total. We monitor the time series of soliton mass and radius in the chain. We remove the first 50 steps to get a flat list of the samples.

\begin{figure}[!h]
\centering
\includegraphics[width=0.5\textwidth]{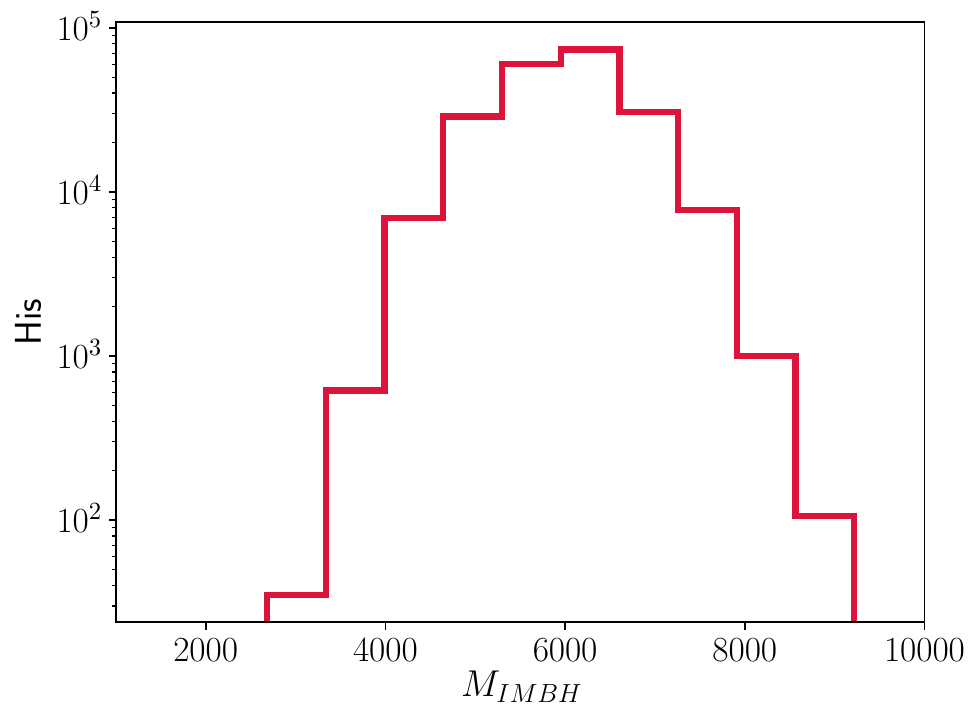}
\caption{Histogram of IMBH from the MCMC fitting. }
\label{Mass-hist}
\end{figure}

Figure \ref{Mass-hist}  presents the histogram of IMBH mass. We have performed an MCMC analysis and allowed a mass to vary in an extended range of $M = (10, 10^6) M_{\odot}$. From the plot it is clear that there is preference of mass around 6000 $M_{\odot}$. More specifically we infer the IMBH mass in the range,

\ba 
\label{IMBH-Mass}
M_{IMBH}  = 5974.13_{- 753.54}^{+752.411} M_{\odot}, 
\ea

%%%%%%%%%%%%%%%%%%%%%%%%%%%%%%%%%%%%%%%%%%%%%%%%%
%%%%%%%%%%%%%%%%%%%%%%%%%%%%%%%%%%%%%%%%%%%%%%%%%

\begin{figure}[!h]
\centering
\includegraphics[width=0.5\textwidth]{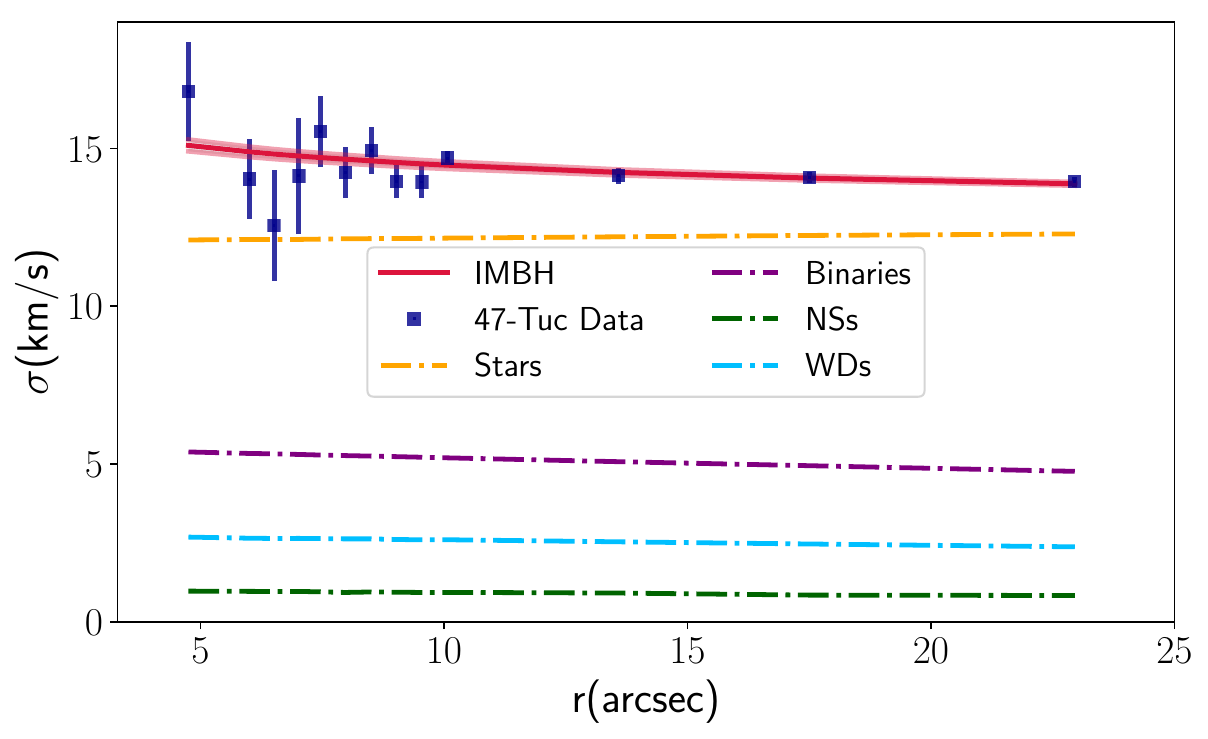}
\caption{Projected velocity dispersion for the case of IMBH. Our theoretical model includes the effect of the IMBH as well as the Stars, Binaries, NSs and WDs. }
\label{vel-fit-IMBH}
\end{figure}

Having estimated the IMBH mass at one sigma level, here we plug in these values in the projected velocity dispersion and visualize it. Figure \ref{vel-fit-IMBH} presents the velocity dispersion for the IMBH case.  Comparing this result with Figure \ref{velocity-projected}  we observe that their velocity profiles are fairly close to each other. This means that at the level of the velocity dispersion IMBH scenario is fairly degenerate with our proposal of soliton. On the other hand, failure to find Radio/Xray from IMBH may still motivate us to use the soliton model as an alternative explanation.

%%%%%%%%%%%%%%%%%%%%%%%%%%%%%%%%%%%%%%%%%%%%%%%%%
%%%%%%%%%%%%%%%%%%%%%%%%%%%%%%%%%%%%%%%%%%%%%%%%%

Finally, we summarize the results of the direct fit for both of the soliton and IMBH models in Table. \ref{tab1}. Here different rows show the number of model parameters, maximum and mean Likelihood, mean mass for the soliton and IMBH, soliton radius and the minimum value of the axion mass.

%%%%%%%%%%%%%%%%%%%%%%%%%%%%%%%%%%%%%%%%%%%%%%%%%%
\begin{table}
\centering
\caption{Direct comparison between the parameters of the soliton model vs IMBH. From up to bottom, different rows show the number of parameters, maximum and mean Likelihood, mean mass of soliton/IMBH, soliton radius and axion mass, respectively. }
\label{tab1}
\begin{tabular}{|lccc|r} 
		\hline
		 $\#$ ~~~~~~~~~~~~~~~~~~~&
		 soliton ~~ & 
        IMBH ~~&
         \\
		\hline
		$N_{Params} $ ~~~ &
		2 &
		1 &
        \\
		\hline
		Max(Log(L))~~~ &
       $-4.26$ & 
        $-4.26$ &
        \\
        \hline
        Mean(Log(L)) ~~~ &
        $-4.92$ & 
        $-4.72$ & 
        \\
        \hline
        Mean(Mass)/$M_{\odot}$ ~~~&
        6189.9 & 
        5974.13&
        \\
        \hline
        Radius/pc ~~~ & 
        $\leqslant  0.27$ & 
        --  &
        \\
        \hline
       Min(m)/eV ~~~ & 
        $ 1.07 \times 10^{-18}$ & 
        --  &
        \\
        \hline
    	\end{tabular}
\end{table}
%%%%%%%%%%%%%%%%%%%%%%%%%%%%%%%%%%%%%%%%%%%%%%%%%%%%

\section{Conclusion}
\label{conc}
In conclusion, we showed that a light scalar field can generate a sufficiently compact dark mass corresponding to an axion of $m \gtrsim 10^{-18} eV$ for a very well studied GC 47-Tuc. We found the parameters of soliton theory including the soliton mass and radius from a direct fit of the projected velocity dispersion with the most recent observational results. This favored a compact soliton profile with a soliton radius $R_s \leq 0.27 pc$. 
We found the fits for case of IMBH and estimated the mass of IMBH at one sigma level. Our results show that these two model are fairly degenerate at the level of the projected velocity dispersion. The soliton idea may however have the advantage over an IMBH interpretation as it does not then conflict with the stringent lack of gas accretion affecting the credibility of the IMBH interpretation. 

Our core size and mass is consistent with Ref.  \cite{Mann:2018xkm}. who concluded that a concentration of stellar BHs with a scale of $\simeq$ 0.045 pc can also in principle explain the rising central velocity dispersion of the visible stars, but the extent to which dynamical friction can gen- erate such a relatively compact core needs to be demonstrated with careful simulations and the loss of such stars from such a dense core by three body encounters must also evaluated. A few such stellar mass BH have been found in deep x-ray observations of GC M22 in Ref. \cite{Strader:2012wj} but not near their centers.

One physical importance of our soliton interpretation is in relation to string theory. Such a $10^{-18}eV$ axion together with the lighter $10^{-22}eV$, as a viable candidate for the dark matter, and much lighter axion, $10^{-33}eV$, to be responsible for the current expansion of the universe, \cite{Kamionkowski:2014zda}, could greatly support the idea of an Axiverse \cite{Arvanitaki:2009fg, Acharya:2010zx, Cicoli:2012sz}, of a discrete mass spectrum of several light axions spanning a wide range of axion mass, generically resulting from higher dimensional compactification. \\

\section*{Acknowledgment}
We are very grateful to John Forbes, Carl-Johan Haster, Abraham Loeb, Michelle Ntampaka, Henry Tye and Francisco Villaescusa for the fruitful discussions.  We especially thanks the anonymous referee for his very constructive comments that improved the paper significantly. R.E. acknowledges the support by the Institute for Theory and Computation at Harvard- Smithsonian Center for Astrophysics as well as Hong Kong University through the CRF Grants of the Government of the Hong Kong SAR under HKUST4/CRF/13.  GFS acknowledges the IAS at HKUST and the Laboratoire APC-PCCP, Universit\'{e} Paris Diderot and Sorbonne Paris Cit\'{e} (DXCACHEXGS) and also the financial support of the UnivEarthS Labex program at Sorbonne Paris Cit\'{e} (ANR-10-LABX-0023 and ANR-11-IDEX-0005-02). TJB thanks IAS hospitality,  SDSS/BOSS data etc. TC acknowledges the grant MOST 103-2112-M-002-020-MY3 of Ministry of Sciences and Technologies, Taiwan. We thank the supercomputer facility at Harvard where most of the analysis was done.

\end{document}